\documentclass[12pt,graphics] {article}
\usepackage[dvips]{graphicx,color}
\newcommand{\bea}{\begin{eqnarray}}
\newcommand{\eea}{\end{eqnarray}}
\newcommand{\be}{\begin{equation}}
\newcommand{\ee}{\end{equation}}

\def\lsim{\mathrel{\lower2.5pt\vbox{\lineskip=0pt\baselineskip=0pt
          \hbox{$<$}\hbox{$\sim$}}}}
\def\gsim{\mathrel{\lower2.5pt\vbox{\lineskip=0pt\baselineskip=0pt
          \hbox{$>$}\hbox{$\sim$}}}}

\begin{document}

\begin{titlepage}
\begin{trivlist}\sffamily\mdseries\large
\item
MTR 01W0000020\\[-0.8ex]
\hrule ~\\[-1.2ex]
{\mdseries MITRE TECHNICAL REPORT}\\[1cm]
\LARGE
\begin{center}
\bfseries
Constraints on Eavesdropping on the\\
BB84 Protocol\\[2.8cm]
\end{center}
\mdseries
\large
G. Gilbert\\[0.8ex]
M. Hamrick\\[0.4ex]
~\\
\large
\textbf{May 2001}\\[3.8cm]
\begingroup\footnotesize
\begin{tabbing}
\textbf{Sponsor:} \phantom{spo} \= The MITRE Corporation \phantom{phantomphantomprospo} \= 
\textbf{Contract No.:} \phantom{pro}\= DAAB07-01-C-C201 \\
\textbf{Dept. No.:} \>W072 \>\textbf{Project No.:} \>51MSR837\\[0.6cm]
The views, opinions and/or findings contained in this report \> \>
   Approved for public release; \\
are those of The MITRE Corporation and should not be \> \>
   distribution unlimited.  \\
construed as an official Government position, policy, or \\
decision, unless designated by other documentation. \\[0.3cm]
\copyright 2001 The MITRE Corporation
\end{tabbing}
\endgroup
~\\
\bfseries
\Large 
MITRE\\
\normalsize
Washington ${\mathbf C^3}$ Center\\
McLean, Virginia\\
\clearpage
\end{trivlist}
\end{titlepage}

\Large
\begin{center}

{\bf Constraints on Eavesdropping on the BB84 Protocol$^{\ast}$}
\normalsize 

\vspace*{45pt}
Gerald Gilbert$^{\dag}$ and Michael Hamrick$^{\ddag}$\\
The MITRE Corporation\\ McLean, Virginia 22102
\end{center}

\begin{abstract}

An undetected eavesdropping attack must produce count rate statistics that are
indistinguishable from those that would arise in the absence of such an attack.
In principle this constraint should force a reduction in the amount of information
available to the eavesdropper. In this paper we illustrate, by considering a particular
class of eavesdropping attacks, how the general analysis of this problem may proceed.

\end{abstract}

\normalsize 
~\\[4.00in]
\hrule width 2.5 in
\vspace*{.175in}
{\footnotesize
$\ast$ This research was supported by MITRE under MITRE Sponsored
Research Grant
51MSR837.\\
\dag ~ggilbert@mitre.org\\
\ddag ~mhamrick@mitre.org
}
\clearpage

\section{Introduction}

The analysis of the types of attacks an eavesdropper (Eve) can make on quantum key
distribution protocols is the subject of intensive research \cite{review, gh_mtr}.
An important
goal of this research is to find upper bounds on the amount of secret key information
that may be compromised by the attack.  In this paper we explore the consequences of the
fact that Eve must conceal her attack from the sender (Alice) and the receiver (Bob) of
the secret key material.  The importance of unperturbed error statistics is well
established in the literature.  However other important signatures of Eve's activity,
such as the effect on overall count rate and the incidence of multiple photon events at
Bob's apparatus, have not yet received sufficient attention.  In this paper we describe
a specific class of attack and analyze the extent to which these constraints can reduce
the amount of information available to an eavesdropper.

\section{Analysis}

Alice and Bob use a BB84 protocol \cite{bb84} augmented
with error correction, authentication,
and privacy amplification.  We consider the special case of weak coherent pulses produced
by attenuating the output of a pulsed laser.  The number of photons in any given pulse is
a Poison-distributed random variable:

\be
\hat\chi\left(\mu,l\right)=e^{-\mu}{\mu^l\over l!}~,
\ee

where $\mu$ is the average number of photons per pulse. The quantum channel is characterized
by a transmittivity $\alpha$, and the quantum efficiency of Bob's detector is $\eta$.
The  dark count
rate of the detector will not be taken into account as it substantially complicates the
analytical results without providing significant additional insight.
Eve begins her attack by measuring the number of photons in the pulse.  This destroys the
coherence of the pulse, but does not affect the polarization, which encodes the information
Alice and Bob intend to share.  The rest of the attack then depends on the number of
photons in the pulse.

If there are 3 or more photons in the pulse, Eve makes a ``direct" attack \cite{gh_mtr}
by making
separate measurements of the polarizations of the photons.  We denote
by ${\hat z}_E\left(l\right)$ the probability
that she is successful in determining the polarization of the photons sent
by Alice \cite{gh_mtr, cb}.
If she is successful, she arranges for an accomplice to inject a photon of the correct
polarization state into the transmission channel near Bob's detector.  This ensures that
any qubit whose value she knows is among those that Bob actually receives.  If the
measurements do not determine the state, she does nothing, and Bob receives no photons
for that pulse.  By itself, this would result in a measurable change in the count
statistics by which Bob could detect the eavesdropping.  Eve compensates for this by
blocking some of the pulses that contain only 1 or 2 photons.  Finally, Eve eavesdrops on
the classical channel during sifting to discover which of the pulses result in bits of
sifted key material.  She then uses her measured polarizations to determine this subset of
Alice and Bob's shared key.

If there are 2 photons in the pulse, Eve blocks the pulse with some probability $p_b$.
Otherwise she makes an ``indirect" attack \cite{gh_mtr}
in which she stores the state of one of the
photons in a quantum memory until after the sifting phase.  By eavesdropping on the sifting
phase, she learns which basis to use to measure the stored photon, thus unambiguously
determining the value of the bit shared by Alice and Bob.  In \cite{gh_mtr}
it is shown that direct
attacks are more effective than indirect attacks when the probability of a photon reaching
Bob is below a certain threshold but that direct attacks are of no use when there are only
2 photons in the pulse \cite{djl}.  Eve chooses to use direct attacks for the stronger
pulses and
indirect attacks for the 2-photon pulses.

Finally, if there is 1 photon in the pulse, Eve blocks the pulse with probability $p_b$.
If the pulse is not blocked, she lets a fraction $1-p_m$ of the pulses pass through.
For the remaining pulses, she picks a basis at random and measures the polarization in
that basis.  She then prepares a photon in the state she has measured and sends it to Bob.
In this case, the photons for which she chose the wrong basis will increase the error rate
seen by Alice and Bob.  We assume that Eve has the technology required to compensate Alice
and Bob's intrinsic error rate (this common assumption constitutes a compromise of
transmission security [2]). She then adjusts the parameter $p_m$ so that the error rate
observed by Alice and Bob is within the limits they expect.

We wish to determine the average amount of information leaked to Eve using this particular
attack.  Consider first the information she obtains from the single-photon pulses.  For
Eve to obtain a bit of information, 1 photon must be in the initial pulse, it must not be
blocked by Eve, it must be measured by Eve, it must reach Bob, it must be detected by his
detector, and all parties must use compatible bases.  If there are 2 photons in the initial
pulse, Eve will obtain a bit of the sifted key if she does not block the pulse, if the
photon she passes is received and detected by Bob, and if Alice and Bob use compatible
bases.  If there are 3 or more photons, Eve's attack is successful if she successfully
measures the state, if Bob detects the photon, and if Alice and Bob use compatible bases.
The resulting expression for the information obtained by Eve is then (the superscript
emphasizes the fact that we do not consider the effect of error correction)

\be
s^{partial}={m\over 2}{\Bigg [}{1\over 2}\chi\left(\mu,1\right)\left(1-p_b\right)p_m
\eta\alpha+\chi\left(\mu,2\right)\left(1-p_b\right)\eta\alpha+z_E\left(\mu\right)\eta
{\Bigg ]}~,
\ee

where we denote by $z_E\left(\mu\right)$ the average of ${\hat z_E}\left(l\right)$ over
the photon number states $l > 3$.

Recall that Eve must tune the parameters $p_b$ and $p_m$ so that the count rate and
error rate
measured by Bob are the same as they would have been with no eavesdropping.  General
expressions for the number of sifted bits per block and the number of errors per block
may be found in \cite{gh_mtr}.
The number of sifted bits in the limit of vanishing dark count is

\be
n={m\over 2}\psi\left(\eta\mu\alpha,1\right)~,
\ee

where $\psi\left(X,l\right)$ is the probability for having $l$ or more photons in a coherent
pulse with mean
photon number $X$.  We next find the number of sifted bits per block when Eve is making
the attack described above:

\be
\hat n={m\over 2}{\Bigg \{}{\Big [}\chi\left(\mu,1\right)+\chi\left(\mu,2\right){\Big ]}
\left(1-p_b\right)\eta\alpha+z_E\left(\mu\right)\eta{\Bigg \}}~.
\ee

We may now determine $p_b$ by requiring $n=\hat n$, which gives

\be
p_b=1-{\psi\left(\eta\mu\alpha,1\right)-z_E\left(\mu\right)\eta\over
{\Big [}\chi\left(\mu,1\right)+\chi\left(\mu,2\right){\Big ]}\eta\alpha}~.
\ee

The value $p_b$ thus obtained must be between $0$ and $1$.  This implies constraints on
the
region of parameter space in which the attack as described can be performed at all.
The consequences of these constraints will be explored in detail in a subsequent
paper \cite{gh}. The principal effect is to force Eve to modify her attack in a way that
further reduces the amount of information she is able to obtain.

Next we consider the error rate.  In the absence of Eve's attack, the number of errors
in Bob's sifted key is

\be
e_T={m\over 2}\psi\left(\eta\mu\alpha,1\right)r_c~,
\ee

where $r_c$ is the intrinsic error probability of the source, channel, and detector
apparatus and we have ignored the dark count.  We assume that Eve can control the
error rate of the channel from Alice to Bob.  If so, then she achieves the strongest
attack by removing the intrinsic errors entirely, in which case we have

\be
{\hat e}_T={m\over 8}\chi\left(\mu,1\right)\left(1-p_b\right)p_m\eta\alpha~.
\ee

Eve therefore adjusts $p_m$ to avoid perturbing the error count, resulting in

\be
p_m={\psi\left(\eta\mu\alpha,1\right)\over \chi\left(\mu,1\right)\eta\alpha}
{4r_c\over 1-p_b}~.
\ee

This quantity must be no larger than $1$, which implies an additional constraint
on the parameter space \cite{gh}.

Note that we have not yet discussed the effect of Eve's attack on the probability
that Bob receives multiple photon counts for a given pulse.  Since this is also a
measurable consequence of her attack, she must adjust her attack to emulate the
unperturbed multiple photon event statistics seen by Bob.  One approach to this
issue has been analyzed in \cite{gh_mtr}. We explore this topic further
in \cite{gh}.

\section{Conclusion}

The Figure shows the information loss, $s^{partial}$ as a function of $\mu$,
for three cases of
interest.  The other parameters are chosen to have reasonable practical
values ($\alpha = 0.01, ~\eta = 0.5, ~r_c  = 0.01$).   The upper curve is
an upper bound for
all attacks consistent with the measured error rate \cite{gh_mtr}.
The lower curve is for
the attack described here, in which Eve adjusts her parameters to match both the
count rate and the error rate.  The curve in the middle is for an attack in which
Eve matches the error rate, but does not attempt to match the count rate.  Note that
the middle and lower curves come close to achieving the maximum for small $\mu$.
Requiring Eve to match the count rate modestly reduces the strength of her attack.

\begin{figure}[htb]
\vbox{
\hfil
\scalebox{0.7}{\rotatebox{0}{\includegraphics{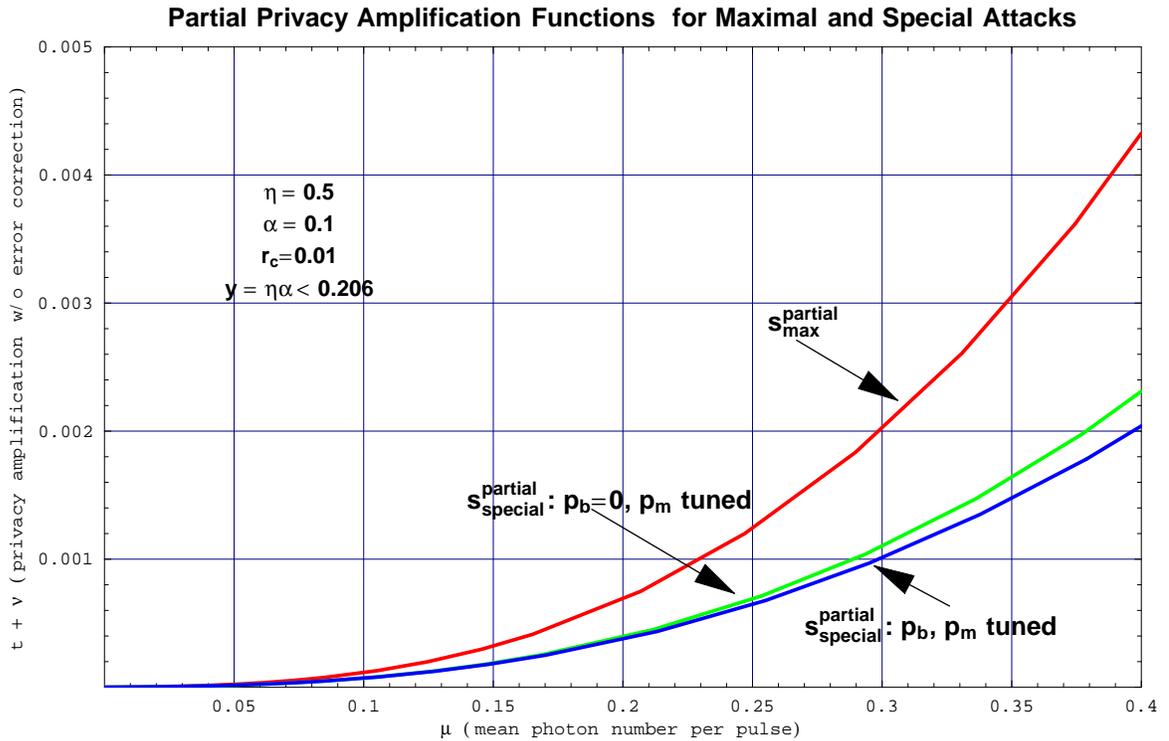}}}
\hfil
\hbox to 0.25in{\ } 
}
\bigskip
\caption{%
Partial Privacy Amplification Functions for Maximal and Special Attacks
}
\label{F:constraint}
\end{figure}

This analysis raises several questions for further research.  First, can these results
be extended to incorporate a general attack by Eve?   Second, are there regions
of ($\mu, ~\alpha, ~\eta, ~r_c$) space for which the enforced reduction of information
available
to Eve through eavesdropping becomes significant? If so, the corresponding rates of
key generation will be greater than current estimates predict.  Finally, what is the
effect of imprecise knowledge of the parameters $\mu, ~\alpha, ~\eta, ~r_c$
on these results?


\end{document}